\documentclass[onecolumn]{emulateapj}

\slugcomment{Draft version}

\shorttitle{Serch for GeV GRBs at the ground}
\shortauthors{Augusto et al.}

\begin{document}

\title{ Search for GeV gamma ray burst signal in the field of view of muon telescopes}
\shorttitle{Search for GeV GRB counterparts at the ground}
\shortauthors{Augusto et al.}

\author{C. R. A. Augusto, V. Kopenkin, C. E. Navia, H. Shigueoka, and K. H. Tsui} 
\affil{Instituto de F\'{\i}sica, Universidade Federal Fluminense, 24210-346, Niter\'{o}i, RJ, Brazil}

\begin{abstract}
We present a search for the muon excess associated in time (preceding or following) with the trigger of gamma ray burst (GRB) detected by the satellite borne instruments.
The 1 Hz rate data was collected at the sea level in the area located at 22S and 43W, inside the South Atlantic Anomaly (SAA) region and close to its center.
Using 1 sec amd 10 sec time binning, we found  the muon excess with  significance $>5\sigma$   
associated with the gamma ray burst GRB091112 observed  by Swift BAT, Fermi GBM and Suzaku-WAM satellites.
The detected muon time profile shows a rough peak  at $\sim T+223$ sec ($5.2\sigma$), and with a duration of 7 sec. 
The FLUKA Monte Carlo simulation of the photon-to-muon conversion in the atmosphere shows that the observed muon excess can be explained by progenitor photons with energies above 10 GeV. 
This energy value is above the limits expected from the maximum synchrotron emission combined with the shock wave evolution. It shows constrains on the external shock synchrotron model for GRB emission in the GeV energy band and favors scenarios with overall afterglow spectrum, ranging from the optical-to-GeV photons.
On the basis of the observed associations with the satellite data, 
we report confidence analysis of the background fluctuations and fluencies in the GeV energy region, that is important for better understanding of the GRB physics.
\end{abstract}


\section{Introduction}
\label{sec1}

Since their discovery by American military satellites (Vela project) in 1967,
 GRBs have evoked intense interest in the scientific community,
 because they are the most energetic explosions in the universe.
 Typical GRBs detected by the satellite borne detectors are constituted by keV to MeV photons
 and they have energy fluxes between $10^{-6}$ to $10^{-4}\; erg\;cm^2$,
 varying in intensity over time scales from milliseconds to several tens of seconds.

\begin{figure}
\epsscale{.80}
\plotone{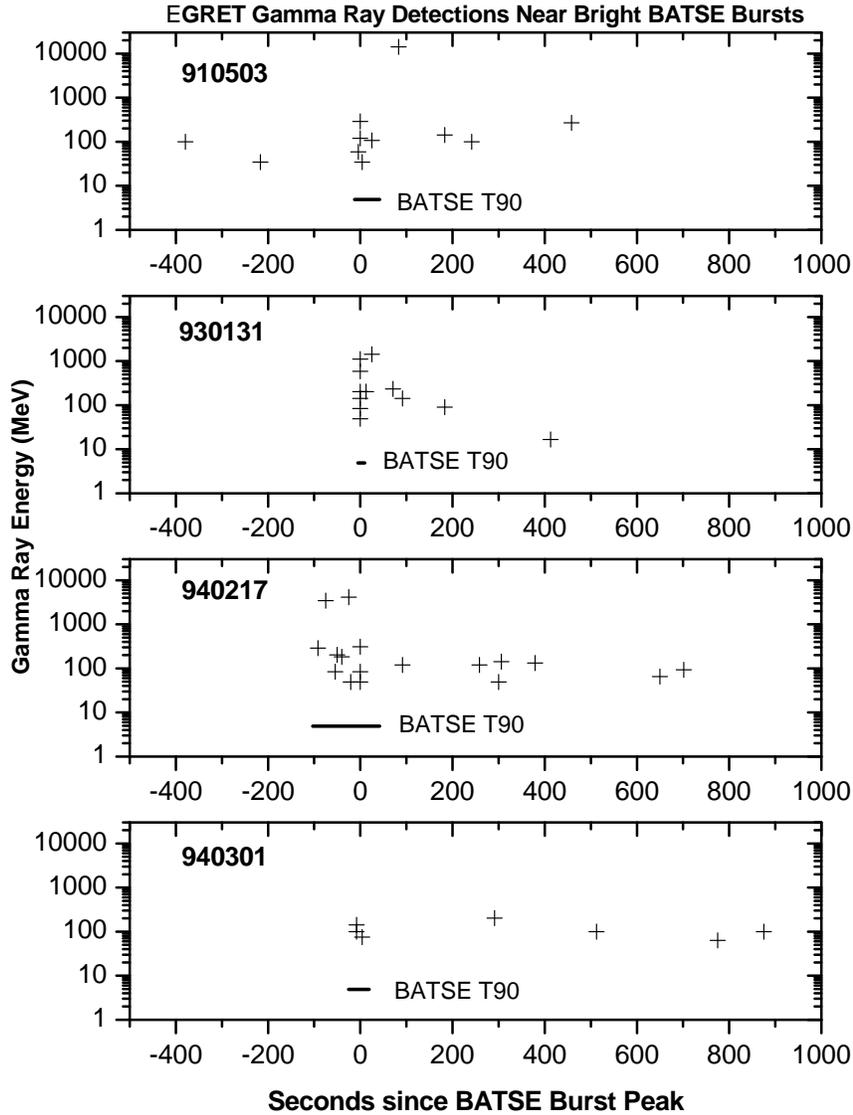}
\vspace*{-1.0cm}
\caption{EGRET detects bright gamma ray BATSE burst. 
The plot shows energy of gamma rays detected by EGRET versus time elapsed since the BATSE trigger. T90 marks the time interval during which 90\% of the sub-MeV signal is detected by BATSE.
\label{fig1}}
\end{figure}

The EGRET instrument installed on board of the Compton Gamma Ray Observatory (CGRO),
 showed that the high energy (MeV-GeV) GRB emission
 lasts longer than the keV emission.
 There were at least five long EGRET GRBs
 detected simultaneously with the bright BATSE sub-MeV emissions
 \citep{sommer94,dingus95,schneid92,hurley94}.
 These MeV-GeV long duration EGRET emissions arrive at the detector discretly, in fragments.
 They are delayed (or anticipated to be delayed) with the respect to the keV-MeV BATSE bursts.
 At present it is not known yet, what is the upper energy limit for 
 this long duration component. 
 Various models \citep{totani99,dermer00,pilla98} 
 predict a fluence in the GeV-TeV range comparable to that in the keV-MeV interval.
 For instance, in case of GRB 940217 \citep{dingus95},
 the Mev-GeV emission persisted for at least $5400s$,
 while the duration of the low-energy (keV) emission lasted for only 180s.
 Fig.~\ref{fig1} presents the overview of this situation.
 
 The long duration GRB event detected by EGRET,
 that shows large time differences in comparison with the BATSE signal,
 can be interpreted as a result of the shock front hitting an interstellar medium, 
 leading to subsequent generation of sub-TeV and TeV gamma rays which in turn
 interact with the infrared and microwave background radiation
 in the interstellar space. 
 As a result, there is a dispersion in the observed arrival time to the detector, because of the differences in the path lengths from the source to the observer.
 
 There is another example of the detection of high energy GRB emissions.
 GRB 080514B was detected by the AGILE gamma-ray satellite \citep{giuliani08}.
 This was the first GRB, after EGRET, where photons above several tens MeV were detected.
 Recently, the Fermi LAT instrument  detected 8 GRBs at energies above 100 MeV \citep{omodei09},
 including extremely energetic GRB080916C (GCN Circular 8246), with the analysis available at \citep{abdo09}).
 This GRB shows that the emission above 100 MeV lasted 20 minutes longer
 than the emission seen by the Fermi GBM in the low energy region.
 In addition to this, the GRB LAT shows that the emission was delayed with respect to the GBM emission.
 
 This signature of the long duration GRBs observed by EGRET and LAT,
 i.e. the large time difference in comparison with BATSE and GBM, respectively,
 suggests that the bulk of the GeV emission arises from an external shock afterglow \citep{kumar09,ghisellini10}. 
If correct, even this scenario is limited to photons with energies below 10 GeV, due to the
limit on the maximum of synchrotron emission, when it is combined with the shock wave evolution \citep{piran10}.

 At the ground level, with the exception of the reported TeV emission from BATSE GRB 970470
 observed by Milagrito (a water Cherenkov experiment and a prototype of Milagro experiment)  \citep{atkins00}, there were no reports of the detected signal in the energy region above 100 GeV for any single GRB.
 For instance, there were approximately 42 satellite-triggered GRBs
 within the field of view of MILAGRO.
 No significant emission was detected from any of these bursts.
 Similar result has been reported by the ARGO project
 (This is a large area detector ($70 \times 70 m^2$)
 installed in Tibet) \citep{vernetto07}.
 
The absence of the high energy gamma rays (GeV to TeV) in the spectrum
 indicates either a high absorption by the extragalactic background light, or 
  atmospheric absorption of air shower particles
 initiated by a gamma ray with energy $\sim 100$ GeV.
 With respect to the second possibility,
 the High Altitude Water Cherenkov (HAWC) project
 represents the next generation of water Cherenkov gamma ray detectors located 
 at the extreme altitudes ($>4$ km asl) \citep{huntemeyer09}, that is 
 in order to minimize the atmospheric absorption.
 
 However, the Tupi muon telescope experiment have produced already some experimental evidence
 for a delayed signal in relation to the Swift GRB 080723
 and the Fermi GBM 081017474 \citep{augusto08c}.
 In addition, the event Konus GRB090315 (GCN Circular 9009) indicated probable correlation with 
 the muon excess. 
 
 The high sensitivity achieved by the Tupi telescopes is in part due to its favorable location.
This detector is situated close to the central region of the South Atlantic Anomaly (SAA) \citep{augusto2011a,augusto10}. The geographical position of this detector allows to achieve the low rigidity threshold for incident primary and secondary charged particles ($>0.8 GeV$). This detector also demonstrated successful detection of high significance muon excess from small solar transient events, 
such as solar flares (the Tupi observation showed association with  GOES (X-rays) and Fermi-BAT (gamma-rays) 
satellites \citep{augusto2011b}), as well as corotating interaction region (CIR)
(the Tupi signal correlated  with the observation by ACE (solar wind) spacecraft detector \citep{augusto2011a}). 
The Tupi telescope detected the smallest solar events observed by the ground level apparatus.
 
This paper is organized as follows:
 Section 2 presents the selection criteria, the search procedure and the result of comparson with the satellite data. A spectral analysis on the basis of the FLUKA Monte 
 Carlo simulation, as well as comparison with other obsrevational results are presented in Sections 3.
 Section 4 is devoted to the confidence analysis, and 
 Section 5 draws conclusions.

\section{Searches and results}
\label{sec2}

\subsection{Selection criteria for the satellite associated signal}
\label{sec2.1}

Some technical details of the Tupi telescopes, such as experimental setup, the analysis of the high detector sensitivity, etc. were published elsewhere \citep{augusto2011a}.

It has been estimated that the field of view of the Tupi telescopes 
allows observations of up to  $\sim 20$ GRBs per year. 
At the same time, the EGRET-BATSE results showed that only long and intense GRBs in the keV-MeV energy range have been associated with the GeV counterparts capable to produce the detected signal at the ground.

The search focussed on the solid angle region along possible GRB arrival
direction, using different time values, with the duration of up to 500 seconds. 
We searched for the muon excess in association with the GRB trigger
given by the satellites (the signals preceding or following the trigger).

At all time the bins for the Tupi data have been tested with a bin selection criterion (BSC). 
The significance of the signal in the i-th bin is defined as following:
 $\sigma_i=(C^{(i)}-B)/\sqrt{B}$,
where $C^{(i)}$ is the measured number of counts, and $B$ is the number of background counts.  
In the absence of any transient signal, $\sigma_i$ follows a Gaussian distribution.

 If there is a muon excess in the time profile (the raw data sequence with one second binning) with a significance above $5\sigma$ during the time interval  
 of up to 500 seconds, around the GRB trigger (indicated by the satellite), then 
 a semi-automatic routine was used to verify the position of the  
  trigger coordinates. If the coordinates were inside of the field of view 
 of the Tupi telescopes, then the signal was verified using 10 seconds binning  with the same applied criteria (above $5\sigma$).
 
 This routine procedure was found to be effective, because in $\sim 97\%$ of cases the
spurious signals arised due to the background fluctuation were eliminated.
 
Furthermore, the detection of the transient events, such as GRB,
  required the axis of the air shower initiated by  photons
  to be located within the field of view of one of the telescopes.
 Thus, the muons  reaching the Tupi telescope will be those that were produced by the most 
  energetic gamma rays (above 10 GeV) of the GRB. 
  Generally, the photo-production of pions (muons) for energies below 10 GeV is considered to be negligible.

The Tupi telescopes operate in the scaler mode.
 In this mode the single hit rate (started since August 2009) is recorded once every second.
 If the Tupi is in the scaler mode, then a transient event, such as a GRB,
 is detected due to photo-muon survivors contained very close to the photon air shower core,
 but not in the entire air shower particles.

 \subsection{GRB 091112}
\label{sec2.2}

 On 12 November 2009, at 17:41:21 UT the Swift Burst Alert Telescope (BAT) triggered and
located the gamma ray burst GRB 091112, trigger=375659
(Mangano, et al., GCN Circ. 10162) and (Palmer, et al., GCN Circ. 10165). 
The trigger was classified as a GRB by the BAT flight software.
In addition to that, the prompt emission of the GRB091112 has been observed  by Fermi GBM ( Briggs, et al., GCN Circ. 10164) and the Suzaku-WAM ( Sugasahara, et al., GCN Circ. 10170). 

\begin{figure}
\epsscale{.80}
\vspace*{-2.0cm}
\plotone{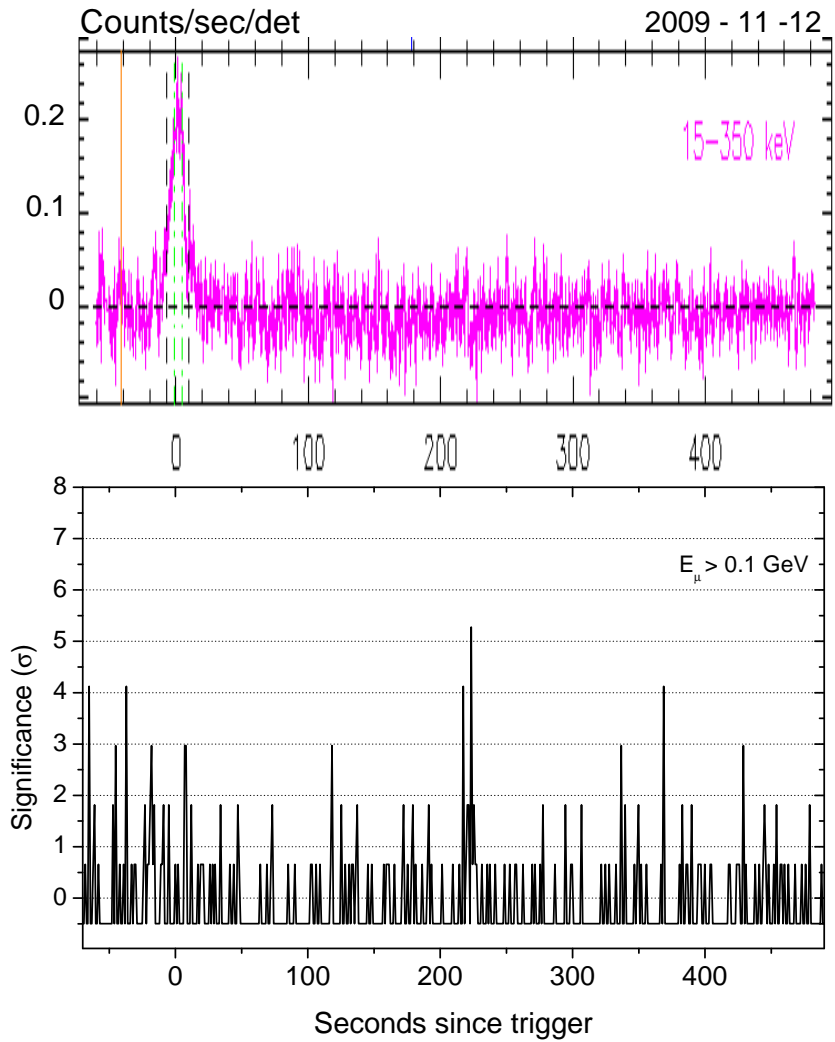}
\vspace*{-1.0cm}
\caption{Comparison between the Swift GRB091112 signal (1 sec bin light curve
 in the  energy range 15-350 keV in top panel)
 and the vertical muon Tupi telescope time profile curve
 with $E_{\mu}>0.1 GeV$ (in bottom panel).
  \label{fig2}}
\end{figure} 

Besides the BAT detector, there were two more detectors on board of the Swift satellite,
XRT (X-ray detector) and UVOT (ultraviolet detector). 
Swift cannot slew to the BAT position, and there are no data in X-ray and UV region for this trigger.

 \begin{figure}
\epsscale{.80}
\vspace*{-2.0cm}
\plotone{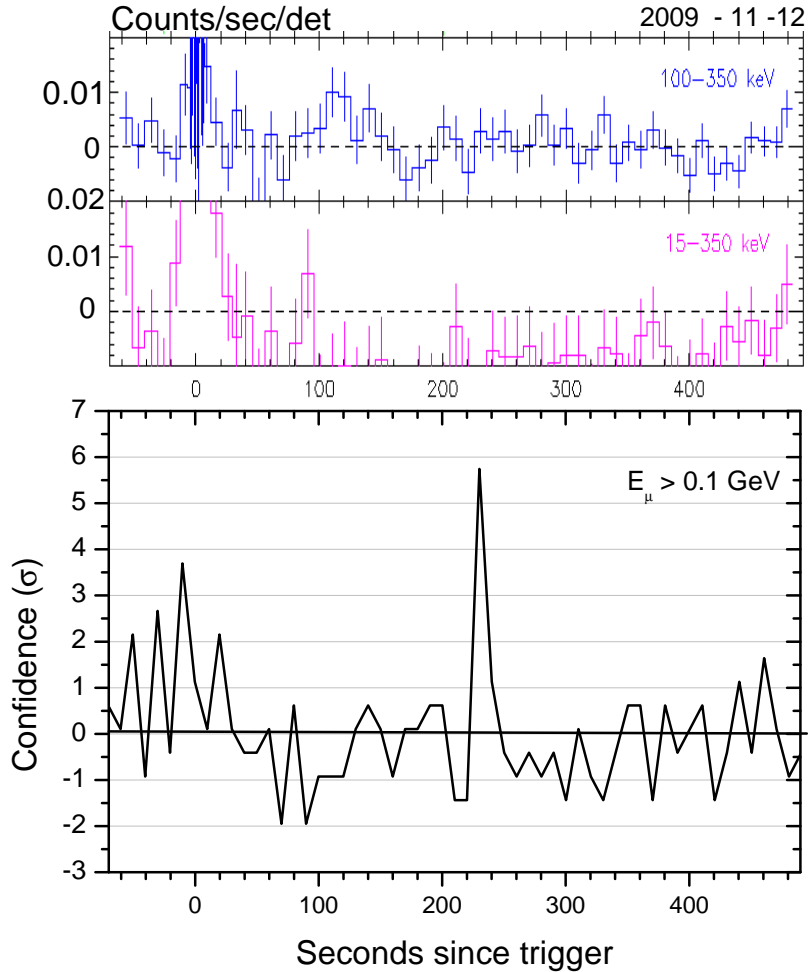}
\vspace*{-1.0cm}
\caption{Comparison of the Swift GRB091112 signal (10 sec binning light curve
 in the two energy intervals, 100-350 keV (top panel) and 15-350 keV (central panel) with
 the vertical muon Tupi telescope  time profile curve
 with $E_{\mu}>0.1 GeV$ (bottom panel).
  \label{fig3}}
\end{figure} 
 
The BAT light curve shows a symmetrical peak starting
at T-15 sec, with maximum at T+2 sec. 
The GRB duration is estimated as $T_{90}=17 \pm 4$ sec,  
with the S/N ratio of 15.19 in the
 (15-350 keV) band. Fig.~\ref{fig2} (top panel) shows the Swift light 
 curve (one sec binning) for the GRB 091112.
 
\begin{figure}
\epsscale{.80}
\vspace*{-0.0cm}
\plotone{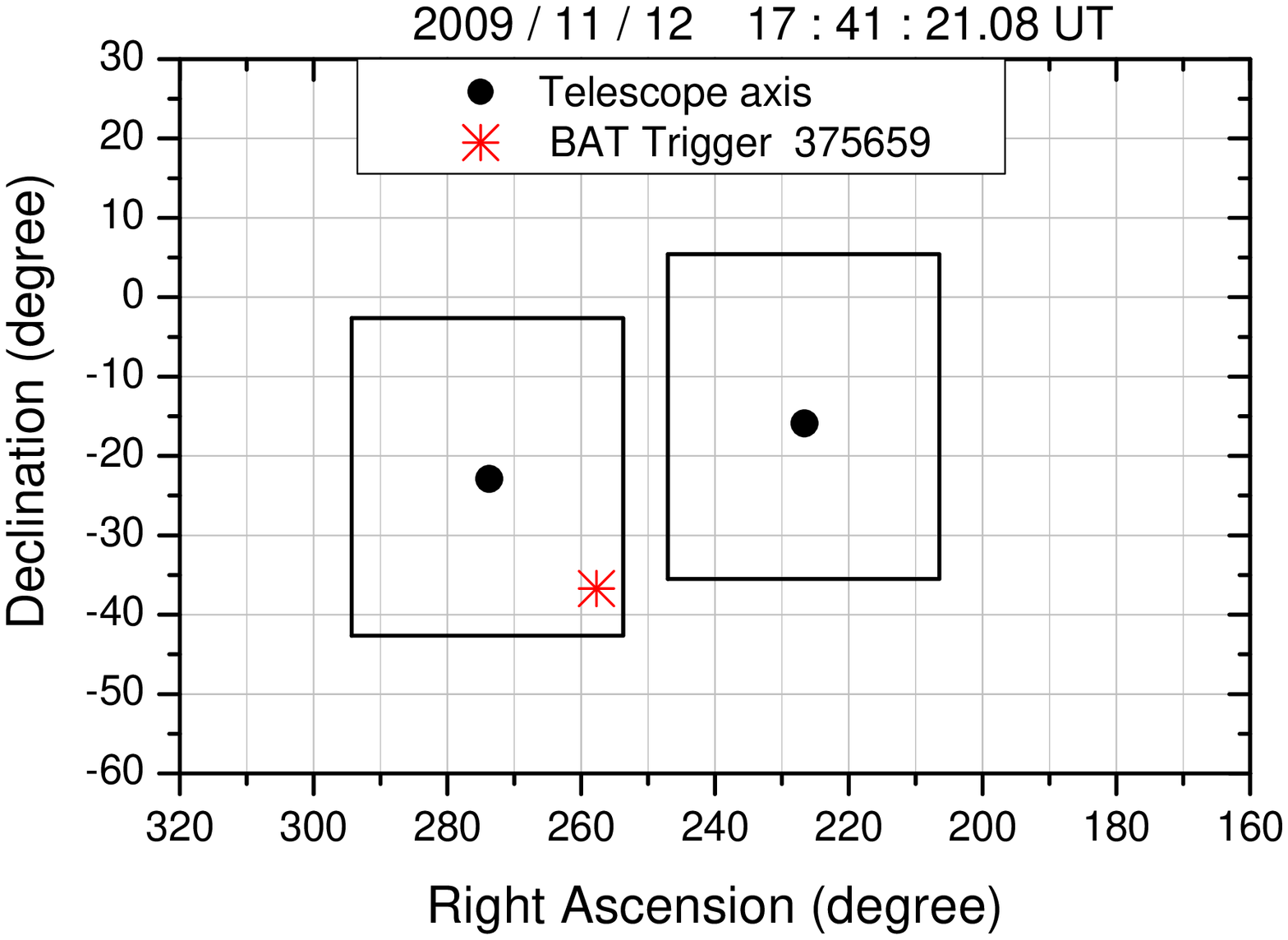}
\vspace*{-8.0cm}
\caption{The equatorial coordinates of  the Tupi telescopes (vertical and inclined) axes (black circles). Squares represent the field of view of the telescopes and the asterisk is the position (coordinates) of the Swift-BAT GRB 091112 trigger.
  \label{fig4}}
\end{figure}

The time profile (raw data) observed by the vertical Tupi telescope and expressed as the muon counting rate (1 second binning) showing several peaks with the significance above $4\sigma$ around the GRB 091112 trigger time is presented in Fig. ~\ref{fig2}(bottom panel).
There is a prominent peak at T+224 sec with the significance of $5.2\sigma$...
According to the selection criteria adopted in this survey, this peak  
satisfies the selection conditions and is considered as true signal.
This signal also appears in the 10 sec binning muon counting rate with the significance $5.4\sigma$, that is above $4\sigma$, as can be seen in Fig. ~\ref{fig3}. 
Comparison between  the Swift (top panel) and Tupi (bottom panel) light curves (10 sec binning) is also presented.
We would like to point out that there is a prompt muon signal at T-7.6 sec.
corresponding to the $T_{90}$ interval detected by the Swift GRB. However, the significance there was only $3.6\sigma$, that is not enough to be considered as a true signal.

In addition to that, the muon excess signal shows  space correlation with the GRB091112. 
This is because during the trigger time the GRB trigger coordinates were within the field of view of the vertical Tupi telescope. 
Fig.~\ref{fig4} shows the  equatorial coordinates together of the telescope axis, together with the Fermi GRB 091112 trigger coordinates.
 The ''squares'' represent the field of view  of the telescopes.

\section{Spectral analysis}
\label{sec3}

\begin{figure}
\vspace*{-5.0cm}
\epsscale{1.00}
\plotone{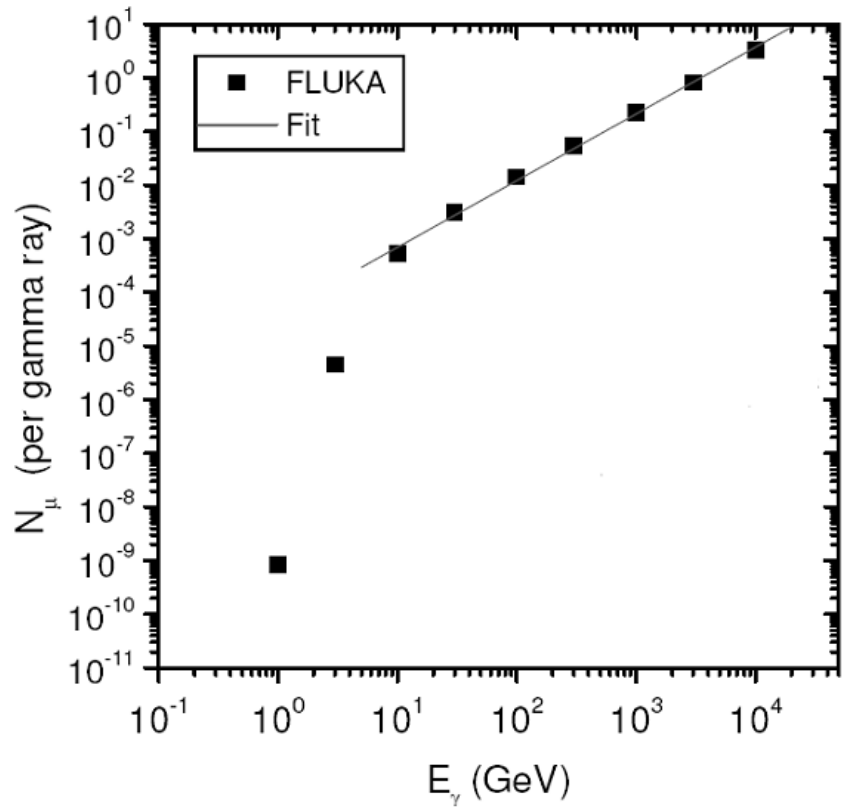}
\vspace*{-5.0cm}
\caption{The number of muons at the sea level per photon, as a function of incident photon energy, from the FLUKA simulation calculations (black squares).  
The calculation is made for a circular area of 10 km radius.  
Black line shows approximation to this function.
\label{fig5}}
\end{figure}
The minimum photon energy needed to produce muons
 with energy $E_{\mu}$ in the atmosphere
 is $E_{\gamma th} \sim 10 \times E_{\mu}$.
 In the Tupi experiment
 the muon energy threshold is estimated as $E_{\mu} \sim 0.1GeV$.
 It means that the minimum proton energy is $E_{\gamma th} \sim 1.0GeV$.
 
 However, Monte Carlo results \cite{poirier02}  showed 
 that an effective photon energy threshold  for muon production in the atmosphere
 is $E_{\gamma th}\sim 10 GeV$ (see Fig.~\ref{fig5}).
 The specific yield function, i.e. the number of muons at sea level per photon,
 as a function of photon energy near the vertical direction, 
 is determined according to the FLUKA Monte Carlos results \cite{poirier02}.
 This FLUKA result can be described by the following fit
 \begin{equation}
 S(E_{\gamma}>10GeV)=A_{\mu} E_{\gamma}^{\nu}exp\left(-(E_{0}/E_{\gamma})^{\lambda} )\right),
 \end{equation}
 where $A_\mu = (6.16 \pm 0.60) \times 10^{-5}$, $\nu=1.183 \pm 0.014$, $E_0=7.13 \pm 0.56$~GeV, $\lambda=1.58 \pm
0.12$.
 
 We assumed that  at the top of the atmosphere
 the energy spectrum of the arrived photons (from GRB) with $E_{\gamma}\geq 10GeV$)
 can be expressed by a single power law function
\begin{equation}
J(E_{\gamma})=A_{\gamma} \left(\frac{E_{\gamma}}{GeV}\right)^{-\beta}.
\label{eq7}
\end{equation}
 The total number of muons  with energies above $E_{\mu}$
 covering an effective area $S_{eff}$ at the sea level, during time T,
 is a convolution between Eq.2 (or Eq.3) and Eq.4,
\begin{equation}
N_{\mu}(\geq E_{\mu})= S_{eff}T\int_{E_{{\gamma}_{min}}}^{\infty}S(E_{\gamma})J(E_{\gamma},T)dE_{\gamma},
\end{equation}
since the muon intensity is given by the integral energy spectrum, and is expressed as
\begin{equation}
I_{\mu}(>E_{\mu})=\frac{N_{\mu}(\geq E_{\mu})}{S_{eff}T }.
\end{equation}
 The muons excess allows us to obtain the coefficient, $A_{\gamma}$,
 for the primary GRB spectrum as
\begin{eqnarray}
A_{\gamma}=I_{\mu}(>E_{\mu}) \frac{1}{A_{\mu}} 
\left(\int_{E_{{\gamma}_{min}}}^{\infty}dE_{\gamma}
\left\{\left(\frac{E_{\gamma}}{\mathrm{GeV}}\right)^{-\beta+\nu}
\exp\left[-\left(\frac{E_{0}}{E_{\gamma}}\right)^{\lambda}\right]\right\}\right)^{-1},
\label{eq10}
\end{eqnarray}

and the integral time fluence can be obtained as 
\begin{equation}
f=T\times \int^{1TeV}_{10GeV} E_{\gamma} 
\left(A_{\gamma} \left(\frac{E_{\gamma}}{GeV}\right)^{-\beta}  \right)
dE_{\gamma}.
\end{equation}

\subsection{Swift GRB091112}
\label{sec3.1}

To explore the possibility that the observed muon excess
 could be explained by the straightforward extension
 of the keV GRB spectrum,
 the power law spectrum is extrapolated to higher energies.
 
\begin{figure}
 \vspace{-1.0cm}
\epsscale{.80}
\plotone{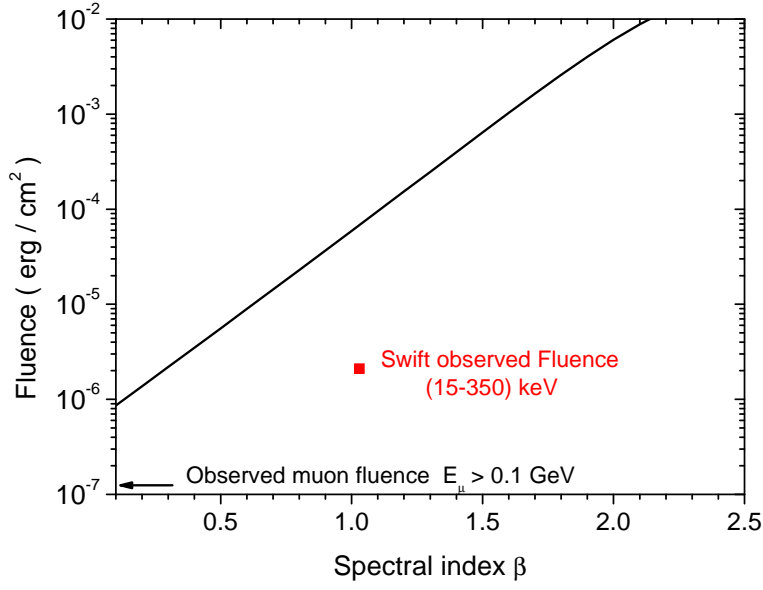}
\vspace{-8.0cm}
\caption{Solid line represents integral time fluence as function of the spectral index, determined on the basis of the Tupi data and FLUKA Monte Carlo simulation and corresponds to the GeV counterpart of the Swift-BAT GRB091112 (red squared). 
The horizontal arrow indicates the observed muon fluence.
  \label{fig6}}
\end{figure} 

\begin{figure}
 \vspace{-2.0cm}
\epsscale{.80}
\plotone{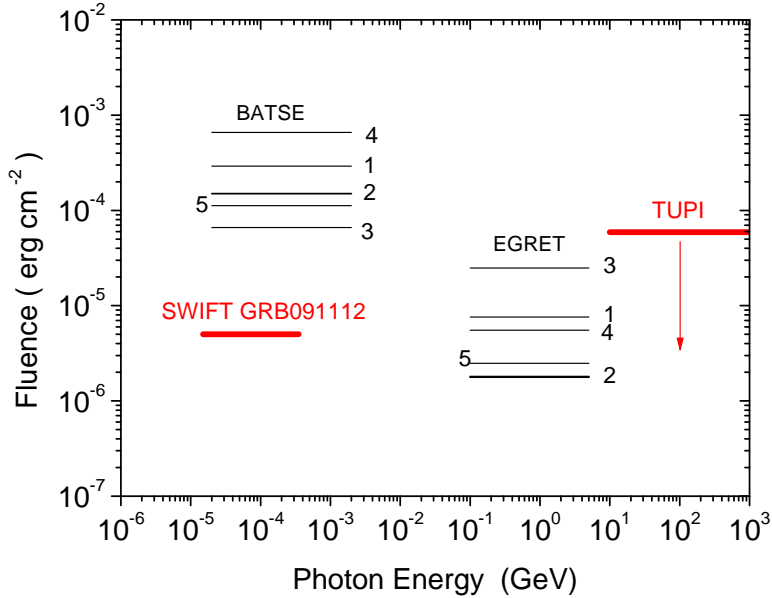}
\vspace{-7.4cm}
\caption{Integral time fluence as a function of the photon energy. 
Black lines represent five EGRET GRBs fluence values  
observed as the high energy counterpart to five GRB BATSE events.
The Tupi fluence of the event presented in this paper corresponds to the Gev counterpart
of the Swift-BAT GRB 091112.
  \label{fig7}}
\end{figure} 

The muon time of flight (ToF) has been studied in the vertical telescope \citep{augusto09}.
The result of the study shows that the detected muons have low energy ($0.1-0.5GeV$). 
The muon background flux in this energy region  strongly depends on the Earth magnetic variations.
The calibration of the muon counting rate to muon flux  should be done at the time close 
 to the muon excess. 
 The muon background flux is estimated as ($9.25 \pm 3.7)\times 10^{-5} muons/cm^2/sec$.
 The muon excess during $7sec$ from $T+223$ to $T+216$,  with a 
 confidence of at least $5.2\sigma$ (see ~\ref{fig2}), is $5.4 \pm 1.8$ higher than the background counting rate, and corresponds to a muon flux of $4.2\pm 1.7 \times 10^{-4} muons/cm^2/sec$.
 
 As it was shown in section ~\ref{sec3}, by using the muon excess and muon intensity
 it is possible to obtain  the photon primary spectrum in the energy region above 10 GeV. 
 However, in order to make a more robust
 comparison with the data and due to the uncertainty in the spectral indices in the high energy region,
the integral time fluence, expressed as a function of the spectral index, was derived. 
Fig. ~\ref{fig6} summarizes the situation. 
If we assume the spectrum for the GeV energy range in the form $N\propto E^{-1}$, such as been observed by Swift for the GRB 091112, for the keV energy range, 
then the fluence in the 10 GeV - 1 TeV energy region is estimated as $5 \times 10^{-5}erg/cm^2$. 
Thus, the fluence in the GeV energy range will be $\sim 10$ times higher than the Swift-BAT fluence (keV range).  

However, we have to note that the high energy spectrum of the observed GRBs is systematically harder (less steep) in comparison with the low  energy spectrum \citep{hurley94}. This means that the above fluence in the GeV energy range, is only an upper limit.

\subsection{Comparison with other observations}  
\label{sec3.2}

\begin{figure}
 \vspace{-1.0cm}
\epsscale{.80}
\plotone{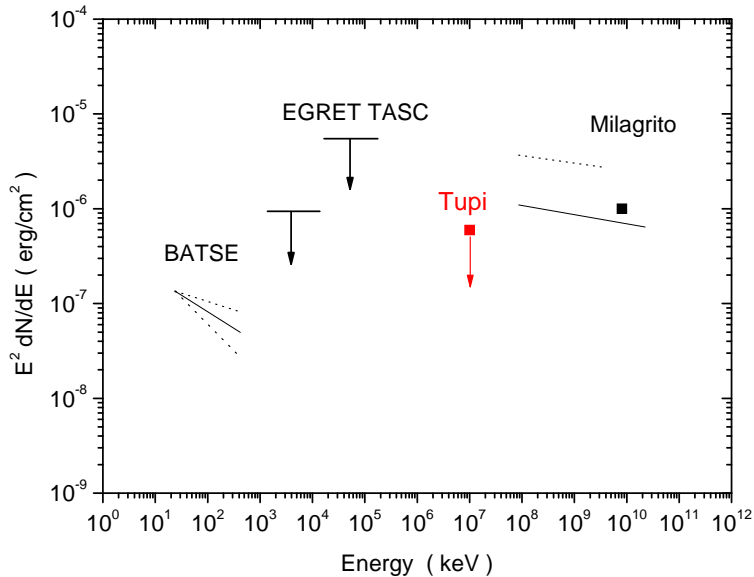}
\vspace{-8.0cm}
\caption{Comparison of the energy distribution of the Tupi event and Milagrito event, that is considered to be a TeV counterpart of the BATSE GRB 970417a. 
This comparison includes the upper limits from the EGRET TASC detector.
  \label{fig8}}
\end{figure} 
 
At first, we present comparison with five long EGRET GRBs detected simultaneously with the bright BATSE sub-MeV emissions. Fig.~\ref{fig7} summarizes the situation. It shows the fluence as a function of the photon energy range. Red lines represent the fluence for the Swift GRB091112 and the associated fluence in
the GeV energy region obtained on the basis of the Tupi-FLUKA result, assuming  straightforward extrapolation 
 of the keV GRB spectrum. By other words,
 the Swift-BAT spectral index is kept constant in the higher energy region. 
 However, if the absolute spectral index in the GeV region is 
 smaller than the absolute spectral index in the keV region, 
 then this result represents an upper limit. The results in Fig.~\ref{fig7} shows a tendency,  the same
 fluence from keV to GeV energy range.

 Black lines are the fluence of the five EGRET events (as BATSE counterpart) 
 observed in the BATSE field of view \citep{li09}.
 This result is consistent with the EGRET/BATSE fluence ratio $0.28$ in average, 
 as well as with the upper limits obtained 
 from a sample of the most luminous GBM bursts with no LAT detection, whose LAT/GBM
  ratio was $0.45$ i the GeV fluence during the first 600 seconds after the trigger \citep{beniamini11}.

These results are in contrast with the Swift-BAT/Tupi fluence ratio obtained here ($>\sim 1$).

We find an average upper limit of LAT/GBM fluence ratio as 0.13 for the GeV fluence.

Second, our comparison is extended to the other ground based observations of GRB. The Milagrito event is considered as the TeV counterpart of the BATSE GRB 970417a \citep{atkins03}. The comparison includes the upper limits at 1 and 10 MeV from the EGRET TASC detector. 
The energy spectrum for these events and the upper limit
of the Tupi events is presented in Fig. ~\ref{fig8}.

\section{Confidence analysis}
\label{sec4}

\begin{figure}
\vspace{-2.0cm}
\epsscale{.80}
\plotone{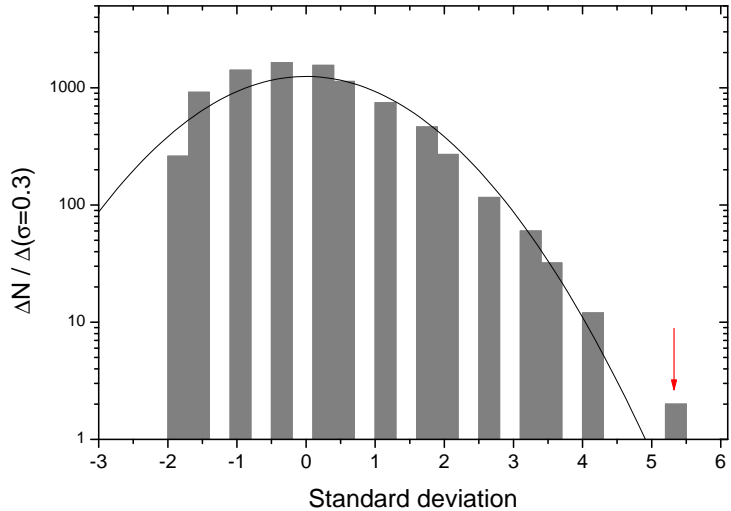}
\vspace{-8.4cm}
\caption{Distribution of the fluctuation counting rate for the Tupi vertical telescope 
(in units of standard deviations). The calculation is made using time windows of $\Delta t=10s$, 
obtained by shifting of the window inside of one hour interval around the Swift-BAT trigger.
\label{fig9}}
\end{figure} 
 To observe the structure of the counting rate (10 sec binning), we performed confidence analysis of the background fluctuations. 
We choose the interval of one hour around 
the trigger time of the Swift GRB 091112. 
 In this time interval the GRB coordinates
 are located within the effective field of view of the Tupi vertical telescope.
 Larger intervals cannot be used,
 since the GRB coordinates would fall outside the field of view
 due to the Earth rotation.
 
 The significance distribution of the  muon counting rate (10 sec binning) in the Tupi vertical telescope is consistent with the Gaussian distribution (see Fig. ~\ref{fig9}). 
 The trials with a confidence of $5.4\sigma$ correspond to the peak at T+224 sec.
 This muon peak shows some structure. 
 One can notice a fast rising exponential decay (FRED) like pulse.
 This behavior is hard to explain as being the background fluctuation.

\section{Conclusions}
\label{sec5}

We searched for the GRB signal in the field of view of the Tupi muon telescopes.
 We found a candidate event with features of likely association  
 with the gamma ray burst GRB091112 observed  by Swift-BAT, Fermi-GBM and Suzaku-WAM.
 
 The GRB 091112 has a fluence of $(5.01\pm 0.74) \times 10^{-6}$ erg cm$^{-2}$, in the 15-350 keV energy band.
 The excess of muons observed by the Tupi vertical telescope associated to this event 
 has a fluence of $\sim 10^{-7}$ erg cm$^{-2}$ ($E_{\mu}>100 MeV$) 
 and corresponds to a 10 GeV-1 TeV photon fluence of up to
 10 times higher than the photon fluence in the keV energy band.
 This value is not far from the observed EGRET fluence in the MeV to GeV energy band.
 
 On the other hand, on the basis of the observed muon excess and muon intensity, we estimated the 
 original GeV photon flux associated with the GRB 091112.
 This estimation is consistent with the arrival of the late photons (about 
 220 seconds later after trigger burst) with energies higher than 10 GeV. 
 This scenario is inconsistent with the synchrotron emission of relativistic 
 electrons accelerated by the external shock model, since there is well known limit (less than 10 GeV) on the maximum synchrotron emission combined with the blast wave evolution. 
 Certainly, these photons might have  different origin, consistent with an overall afterglow scenario, 
(for photons from optical-to-GeV band), when the low magnetic field in the emitting region quenches the lower energy emission.

 Here we summarize several favorable characteristics which allow us to suggest
 that the analyzed muon excesse was indeed the GeV counterpart of the Swift-BAT GRB091112.
  First, the index of the GRB 091112 energy spectrum (the slope) is small. This factor is extremely favorable,  
 since the spectrum has been extrapolated to higher energies, where the photon intensity  
 does  not have a drastic reduction.   
 Second,  the muon signal  at $T+220s$ from the GRB trigger time shows  high significance (above $5\sigma$) in the raw data (1 sec binning) and it persists with a significance above $5\sigma$ for 10 sec binning.  
 Thus, this signal shows a clear deviation from the statistical background  behavior. 
 Third, the GRB 091112 trigger coordinates were within the field of view of the vertical Tupi telescope (telescope pointing to the zenith). This means that the muon absorption in the atmosphere was the smallest.

Finally, considering the photon absorption by the extragalactic 
background light and using the fluence value estimated here in the energy range 10-1000 GeV, 
we estimate the  upper limit on the redshift of the GRB 091112 as  $z\leq 1.0$.

\acknowledgments

   This work is supported by the National Council of Research (CNPq) of Brazil,
 under Grant $479813/2004-3$ and $476498/2007-4$.
 We are grateful to various catalogs available on the web
 and to their open data policy, especially to the GCN report and the Swift batgrbproduct Analysis.

{}

\end{document}